\documentclass[aps,amsfonts,pra,twocolumn,showpacs,superscriptaddress]{revtex4-1}
\usepackage{epsfig,amsmath,amssymb,bm,epsf,graphicx,psfrag,color,dsfont,natbib,caption,subcaption,epstopdf,multirow,diagbox}
\usepackage{bbold}
\usepackage{float}

\makeatletter
     \renewcommand\@make@capt@title[2]{%
      \@ifx@empty\float@link{\@firstofone}{\expandafter\href\expandafter{\float@link}}%
       {\textbf{#1}}\@caption@fignum@sep#2\quad}%
     \makeatother
 
\makeatletter 
\renewcommand{\fnum@figure}{\textbf{Figure~\thefigure}}
\makeatother

\newcommand{\bc}{\begin{center}}
\newcommand{\ec}{\end{center}}
\newcommand{\beq}{\begin{equation}}
\newcommand{\eeq}{\end{equation}}

\def\ket#1{\vert#1\rangle}
\def\bra#1{\langle#1\vert}

\newcommand{\iden}{\mathds{1}}

\begin{document}

\title{Detector Tomography on IBM 5-qubit Quantum Computers and Mitigation of Imperfect Measurement}

\author{Yanzhu Chen}
\affiliation{C. N. Yang Institute for Theoretical Physics, State University of New York at Stony Brook, Stony Brook, NY 11794-3840, USA}
\affiliation{Department of Physics and Astronomy, State University of New York at Stony Brook, Stony Brook, NY 11794-3800, USA}
\author{Maziar Farahzad}
\affiliation{Department of Physics and Astronomy, State University of New York at Stony Brook, Stony Brook, NY 11794-3800, USA}
\author{Shinjae Yoo}
\affiliation{Computational Science Initiative, Brookhaven National Laboratory, PO Box 5000, Upton, NY 11973-5000, USA}
\author{Tzu-Chieh Wei}
\affiliation{C. N. Yang Institute for Theoretical Physics, State University of New York at Stony Brook, Stony Brook, NY 11794-3840, USA}
\affiliation{Department of Physics and Astronomy, State University of New York at Stony Brook, Stony Brook, NY 11794-3800, USA}
\affiliation{Institute for Advanced Compuational Science, State University of New York at Stony Brook, Stony Brook, NY 11794-5250, USA}

\date{\today}

\begin{abstract}
We use quantum detector tomography to characterize the qubit readout in terms of measurement POVMs on IBM Quantum Computers IBM Q 5 Tenerife and IBM Q 5 Yorktown. Our results suggest that the characterized detector model deviates from the ideal projectors by a few percent. Further improvement on this characterization can be made by adopting two- or more-qubit detector models instead of independent single-qubit detectors for all the qubits in one device. An unexpected behavior was seen in the physical qubit labelled as qubit 3 of IBM Q 5 Tenerife, which can be a consequence of detector crosstalk or qubit operations influencing each other and requires further investigation.  This peculiar behavior is consistent with characterization from the more sophisticated approach of the gate set tomography. We also discuss how the characterized detectors' POVM, despite deviation from the ideal projectors, can be used to estimate the ideal detection distribution.
\end{abstract}

\maketitle

\section{Introduction}

	Recent manufacture of 49, 50 and 72 superconducting qubits from companies such as Intel, IBM and Google gives prospect   of demonstrating quantum advantage in not distant future. However, these and near-future machines  are at best noisy intermediate-scale quantum (NISQ) processors~\cite{NISQ}. Therefore, developing and harnessing tools for characterizing noise and error, mitigating them, and verifying quantum processing will be essential in running programs on quantum devices and in further locating the parameter windows in applications towards quantum advantage. Several tools have been developed, including tomographic ones for {\it quantum states and processes}~\cite{NielsenChuang,Leonhardt95,Hradil97,JamesKwiatMunroEtAl01,TurchetteHoodLangeEtAl95,NielsenChuang97,PoyatosCiracZoller97}. But these rely on accurate measurement and/or state preparation, for which the system may not have, and the methods do not scale favorably with the system size. If one is only concerned with partial characterization, such as the average gate error rate, then the so-called {\it randomized benchmarking}~\cite{EmersonAlickiZyczkowski05,KnillLeibfriedReichieEtAl08,MagesanGambettaEmerson11} provides a reliable estimation independent of state preparation and measurement error. Although these tools seem to be standard, there are still some aspects of them not fully explored. 

	When one speaks of qubit decoherence, there are typically two associated processes: (1) relaxation (with time called $T_1$),  usually related  to the transition of the excited state(s) back to the ground state or the system returning to thermal equilibrium and (2) dephasing (with time $T_2$), related to  off-diagonal elements of the density matrix decaying exponentially with time.  In reality, a qubit will couple to the environment and such interaction  (and with other qubits in an undesired way) will induce relaxation and dephasing, and possibly other ways causing decoherence. These will be loosely referred to as  noise, and any quantum gate that does not operate as desired is said to have errors.    For instance in the IBM quantum computers, the error rate in measurement readout (2-10\%) is comparable to that of two-qubit gates (3-7\%) and both rates are greater than that of single-qubit gates (0.1-0.2\%) by one order of magnitude. Single-qubit state preparation for short circuits is to some extent of high fidelity, but the computation for longer circuits will inevitably suffer from noise. One tomographic tool that sometimes gets overlooked is the so-called {\it quantum detector tomography}~\cite{Fiurasek2001}, more recently discussed in photon detectors~\cite{LundeenFeitoColdenstrodt-RongeEtAl09,FeitoLundeenColdenstrodt-Ronge09,RenemaFruccideDoodEtAl12,RenemaGaudioWangEtAl14}, which seems to provide a first tool to improve the readout or detector characterization, via short quantum circuits involving single-qubit gates. Because of the measurement error is higher than state preparation (of $|0\rangle$) and single-qubit gates on IBM Q devices, we perform  the quantum detector tomography to characterize the detectors. We point out some behavior revealed by experiments that require further investigation into physical devices, beyond the setting of quantum circuits. 
	
	We remark that a more thorough characterization scheme that makes the fewest assumptions is {\it gate set tomography}~\cite{Blume-Kohut2013, Blume-Kohut2017}, where an initial state, a set of quantum gates, and a positive-operator valued measure (POVM) are characterized simultaneously. Since this requires a large number of gate sequences, some of which are very long, it is currently limited to single-qubit and two-qubit processes in practice. Another recently proposed scheme that is less costly tries to characterize state preparation and measurement iteratively~\cite{Keith2018}.
	
	In the next section (\ref{sec:main}) we briefly review a few tomographic tools, and emphasize that for detectors. We present our experimental results in Sec.~\ref{sec:results}. An unexpected behavior was seen in the physical qubit labelled as qubit 3 of IBM Q 5 Tenerife. Its detector characterization seems to be different when it is done alone (with other qubits being idle) from when it is done when other qubits are also in operation. This can be a consequence of detector crosstalk or qubit operations influencing each other and requires further investigation. In Sec.~\ref{sec:ideal}, we describe how such characterized readout can be used for mitigation of measurement error, in the sense of inferring ideal measurement statistics. In Sec.~\ref{sec:GST}, we also use the gate set tomography and compare its detector characterization with that from the simple detector tomography. We make concluding remarks in Sec.~\ref{sec:conclude}. Some experimental data and further results of GST are presented in the Appendix, including QDT for the 14 qubits of IBM Q 16 Melbourne.

\section{Tomographic tools for states, detectors and processes} \label{sec:main}

	\smallskip \noindent\textbf{Quantum State Tomography} (QST). The idea of quantum state tomography was proposed~\cite{VogelRisken89,Leonhardt95} early in the context of quantum optics using quasiprobability distributions, and it has become a standard procedure in measuring multiple qubits~\cite{Hradil97,JamesKwiatMunroEtAl01}.
The basic idea is that one has a set of projectors or more general POVM elements $\{\Pi^{(i)}\}$ (e.g. ${\cal A}_1\equiv\{|0/1\rangle\langle 0/1|, |+/-\rangle\langle+/-|, |\pm i\rangle\langle\pm i|\}$, corresponding to eigenstates of Pauli matrices), and one measures them with respect to an unknown state $\rho$, yielding a set of data $p_{\rho,i}={\rm Tr}(\rho \Pi^{(i)})$. Since $\{\Pi^{(i)}\}$ is chosen to be (over-)complete, one can from the statistics $p_{\rho,i}$'s infer the best estimate  $\tilde{\rho}$, via e.g. the maximum likelihood method (MLE)~\cite{JamesKwiatMunroEtAl01}. The approach was later extended to a `hedged' version~\cite{Blume-Kohout10} and a mean Bayesian version~\cite{Blume-KohoutNJP10} that deal with certain drawbacks of MLE~\cite{FerrieBlume-Kohout18,SchwemmberKnipsRichart15}. However, QST requires  ${\cal O}(3^n)$  different measurement bases for $n$ qubits, but compressed sensing can be used to ameliorate this~\cite{GrossLiuFlammiaEtAl10}. 

	Nevertheless, quantum state tomography remains an indispensable ingredient in characterizing small quantum systems, and even a partial tomography (for some part of a larger system) can also be useful when one is verifying some properties, such as the existence of entanglement,  that may not require a complete global wavefunction.  \textit{However, most of the description relies on the assumption that  almost perfect projective, von Neumann measurements can be performed}. Here, we will consider a more realistic scenario where measurements are not necessarily projective, as in e.g.  For example, IBM quantum computers whose measurement errors are not negligible, of order 2\% to 5\%. (See  manufactures' released data for devices' properties, e.g. on IBM Q Experience or Rigetti Computing, but some useful information was listed in the Appendix of Ref.~\cite{PokharelAnandFortmanEtAl18}). 

	\smallskip \noindent\textbf{Quantum Process Tomography} (QPT). Related to state tomography is the characterization of a quantum process, which may arise from application of a gate or evolution of a system that possibly couples to its environment. In the later case, it is commonly considered in the Markovian limit, and one arrives at the so-called master equation for the system state $\rho(t)$~\cite{NielsenChuang},
	\begin{widetext}
	\beq
		\frac{d\rho(t)}{dt}=-\frac{i}{\hbar}[H,\rho]+{\cal L}(\rho)=-\frac{i}{\hbar}[H,\rho]+{\sum}_j \left(2L_j\rho L_j^\dagger -\{L_j^\dagger L_j,\rho\}\right),
		\label{eqn:Lindblad}
	\eeq
	\end{widetext}
	where $H$ is the system Hamiltonian, and $L_j$'s are the Lindblad operators, representing the effect of coupling to environment. One can describe the change of $\rho$ in a discrete time step $\Delta t$ as a quantum process, $\rho(t_0)\rightarrow \rho(t_0+\Delta t)={\cal E}(\rho)$. A general quantum process ${\cal E}$ can be described by a set of Kraus operators $E_j$, so that its action on $\rho$ is ${\cal E}(\rho) = \sum_j E_j \rho E_j^{\dagger}$, where without loss of generality we can assume ${\cal E}$  to be trace preserving: $\sum_j E_j^\dagger E_j=I$, unless there is some loss or leakage. The procedure to infer ${\cal E}$ is called quantum process tomography, which is natural to consider given QST~\cite{TurchetteHoodLangeEtAl95,NielsenChuang97,PoyatosCiracZoller97,NielsenChuang}. It is possible to infer the quantum process because of linearity and if one applies (unkonwn) ${\cal E}$ to a complete basis of a density matrix, e.g. $|k\rangle\langle l|\rightarrow {\cal E}(|k\rangle\langle l|)$, then by measuring the output the process can be determined~\cite{NielsenChuang}. The matrix element $|k\rangle\langle l|$ can be expressed in terms of a linear combination of different states $|\psi\rangle\psi|$ in, e.g.  $	{\cal A}_1\equiv\{|0/1\rangle\langle 0/1|, |\!+\!/\!-\rangle\langle+\!/\!-\!|, |\pm i\rangle\langle\pm i|\}$ for one qubit,  and thus \textit{quantum process tomography uses quantum state tomography as a sub-routine.} Instead of varying the input states over some `complete' (or even over-complete) set, such as ${\cal A}_1$ above, one can also use a bi-partite maximally entangled state $|\Psi\rangle_{AB}$, where the party $A$ corresponds to the system that will be acted on by the process ${\cal E}_A$, and the party $B$ acts as an ancillary role. Then  
 state tomography on the resulting bi-partite system (after $A$ undergoing the process ${\cal E}$) gives identical determination of the process ${\cal E}$~\cite{AltepeterBranningJeffreyEtAl03}.

However, in currently available small-scale quantum computers, both measurement and state preparation have errors, and if there is some separation of rates in these different types of errors, as in IBM and Rigetti quantum computers, then we can give better individual characterization.

	\smallskip \noindent\textbf{Quantum Detector Tomography (QDT)}. The discussions above point to the importance of a third, often ignored, tomography for detectors~\cite{Fiurasek2001}, perhaps due to the recent focus on photon detectors~\cite{LundeenFeitoColdenstrodt-RongeEtAl09,FeitoLundeenColdenstrodt-Ronge09,RenemaFruccideDoodEtAl12,RenemaGaudioWangEtAl14}. But we emphasize that in order to ascertain results of computation, detector characterization is as important as state preparation and gate operation.

	In the state tomography we introduce measurement probabilities $p_{\rho,i}={\rm Tr}(\rho \Pi^{(i)})$ for estimating unknown $\rho$ with known measurement operators $\Pi^{(i)}$'s. The detector tomography is a dual viewpoint: with a set of known states $\{\rho_f\}$, one is asked to estimate a fixed but unknown set of measurement operators $\{\Pi^{(i)}\}$ characterizing a detector. Here we formulate qubit detectors that are most relevant to realistic measurement in cloud quantum computers, such as IBM Q and Rigetti's. The usual assumption is that the set of states $\{\rho_f\}$'s is well known or at least with much smaller error rates than detection. For  the state preparation in $|0\rangle$, the typical ground state of superconducting qubits, it is fairly accurate. Moreover, in these systems the single-qubit gates have higher fidelity (than the measurement and two-qubit gates), and only $Z$ measurement can be implemented. Measurement in other bases needs to be actively made by the users to perform a suitable rotation before the $Z$ measurement. Hence we will consider two measurement operators $\Pi^{(0)}$ and $\Pi^{(1)}$ for a single qubit, which is constrained by the trace-preseving condition that $\Pi_0+\Pi_1=\iden$. In the ideal case, $|0\rangle\langle0|=(\iden+\sigma_3)/2$ and $|1\rangle\langle1|=(\iden-\sigma_3)/2$.

	Let us denote for convenience,
	\beq
		\Pi_1^{(n)}=\sum_{i=0}^3a_i^{(n)}\sigma_i,
	\eeq
	where the subscript of $\Pi_1^{(n)}$ means single-qubit detector and $(n)$ denotes the measurement outcome 0 or 1. The Pauli basis is $\sigma_0=\iden, \sigma_1=\sigma_x, \sigma_2=\sigma_y, \sigma_3=\sigma_z$. We can use a vector $\vec{a}^{(n)}=(a_0^{(n)}, a_1^{(n)}, a_2^{(n)}, a_3^{(n)})$ to collectively denote the parameters. There are some constraints: (1) $\vec{a}^{(0)}+\vec{a}^{(1)}=(1,0,0,0)$, and (2) $|a_0^{(n)}|^2 \ge \sum_{i=1}^3 |a_i^{(n)}|^2$ in order for $\Pi^{(n)}$ to be non-negative. We choose and prepare $\rho$ from the 6-element set ${\cal A}_1$ (listed above), and those other than $|0\rangle$ can be prepared from it with relatively high fidelity by single-qubit gates. Then the measurement process accumulates a set of data $P_{\rho_i,n}={\rm Tr}(\rho_i \Pi_n)$, which is a $6\times2$ matrix for each detector. From this we can find the best fit, under the above constraints, to extract $\vec{a}^{(n)}$ that describes the action of the detector. In adopting this model, we have made the assumption that there is no crosstalk between the qubits in one device so that the detectors are viewed as independent. Relaxing this assumption a little bit, we can have a multi-qubit detector model, where a binary string is produced as the measurement result. Just like the single-qubit case, the $N$-qubit detector model is written as:
	\beq
		\Pi_N^{(\vec{n})} = \sum_{\vec{i}} c^{(\vec{n})}_{\vec{i}} \sigma_{i_0} \otimes ... \otimes \sigma_{i_j} ... \otimes \sigma_{i_{N-1}},
		\label{eq:multiqubit}
	\eeq 
	where the binary string $\vec{n}=(n_0, ..., n_{N-1})$ is the measurement outcome and each component of $\vec{i}=(i_0, ..., i_{N-1})$ runs from $0$ to $3$. It is natural to ask whether measuring only one single qubit in the device gives the same result as measuring all the qubits and tracing out the other irrelevant ones. This question will be addressed in our experiments.
	
	In an experiment, the characterized detectors can be used to perform QST on the resultant state, hence mitigating the effect from detector errors. However, we remark that some correction can be made even without a set of informationally complete set of measurements on the state; see discussions below in Sec.~\ref{sec:ideal}.
	
	\subsection{Maximum likelihood estimation}

	In this section we summarize the MLE analysis for detectors that we will employ~\cite{Fiurasek2001}. The log likelihood function is defined as:
	\beq
		{\rm log}\mathcal{L} = \sum_{n} \sum_{i} f_{n,i} {\rm log} {\rm Tr} \big( \Pi^{(n)} \rho_i \big),
	\eeq
	where $\{\Pi^{(n)}\}$ is the POVM characterizing the detector and $f_{n,i}$ is the frequency of measuring the state $\rho_i$ and obtaining outcome $n$. The sum over index $i$ contains an informationally complete set of test states. The normalization constraint $\sum_{n}\Pi^{(n)}=\iden$ is implemented by Lagrange multipliers. Maximization with the constraint leads to the equation:
	\beq
		\Pi^{(n)} = R^{(n)} \Pi^{(n)} R^{\dagger (n)}. \label{eq:iter1}
	\eeq
	$R^{(n)}$ is determined by the normalization constraint, and is given by:
	\beq
		R^{(n)} = \sum_i \frac{f_{n,i}}{p_{n,i}} \big( \sum_m\sum_{j,k} \frac{f_{m,j}f_{m,k}}{p_{m,j}p_{m,k}} \rho_j \Pi^{(m)} \rho_k \big)^{-\frac{1}{2}} \rho_i,  \label{eq:iter2}
	\eeq
	where $p_{m,j}$ denotes the theoretical probability of measuring the state $\rho_j$ and obtaining outcome $m$. Note that $R^{(n)}$ is a function of the POVM $\{\Pi^{(m)}\}$, not only through the explicit dependence, but also because $p_{m,j}={\rm Tr} \big( \Pi^{(m)} \rho_j \big)$. In our analysis, we choose the multi-qubit Pauli matrices as the basis to express $\{\Pi^{(m)}\}$ and $\{\rho_j\}$. Each iteration starts with updating $\{\Pi^{(m)}\}$ according to Eq.~\ref{eq:iter1}, and ends with calculating $\{R^{(m)}\}$ from Eq.~\ref{eq:iter2} for the next iteration. The termination condition is set as:
	\beq
		\sum_n || \Pi^{(n)}_t-\Pi^{(n)}_{t+1} || < \epsilon,
	\eeq
	where the subscript denotes the $t$-th and the $(t+1)$-th iterations, the norm is taken to be Frobenius norm, and $\epsilon$ is some (arbitrarily chosen) cutoff value. Positivity and normalization are preserved as long as the initial values of $\{\Pi^{(m)}\}$ form a POVM. It is worth mentioning that $\epsilon$ should be sufficiently small such that the numerical error introduced by this cutoff would be smaller than the uncertainty in the estimated parameters due to statistical fluctuations. 

\section{Results of quantum detector tomography} \label{sec:results}

	\begin{figure*}[t]
		\centering
		\begin{subfigure}[b]{0.8\textwidth}
			\includegraphics[width=120mm,height=24mm]{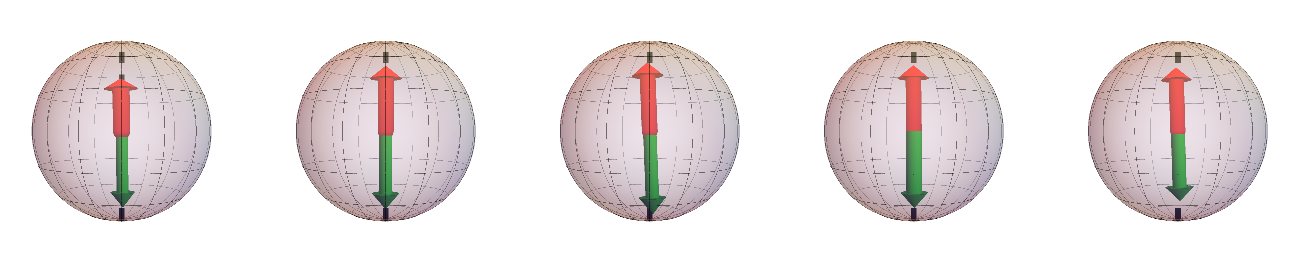}
			\caption{Individual measurement on IBM Q 5 Tenerife.}
		\end{subfigure}
		
		\begin{subfigure}[b]{0.8\textwidth}
			\includegraphics[width=120mm,height=24mm]{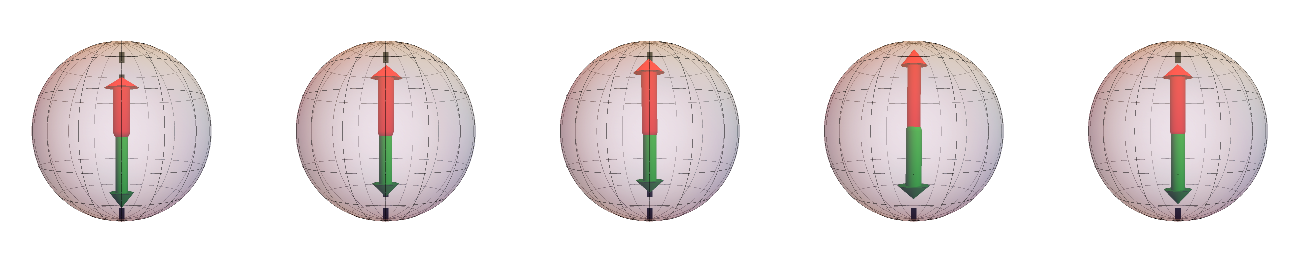}
			\caption{Parallel measurement on IBM Q 5 Tenerife.}
		\end{subfigure}
		
		\begin{subfigure}[b]{0.8\textwidth}
			\includegraphics[width=120mm,height=24mm]{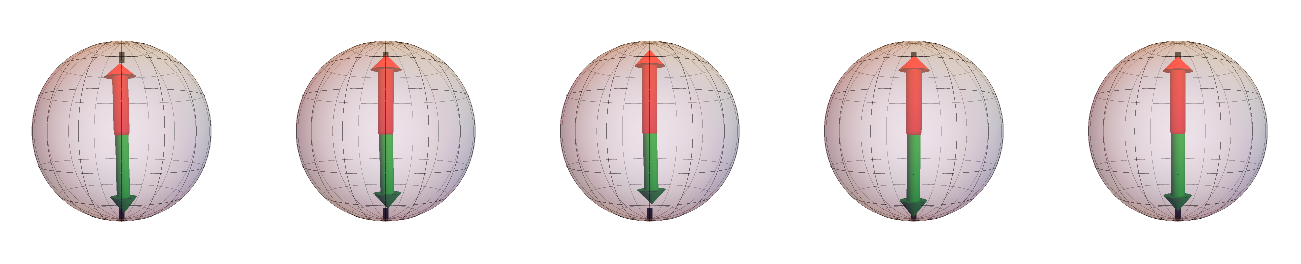}
			\caption{Individual measurement on IBM Q 5 Yorktown.}
		\end{subfigure}
		
		\begin{subfigure}[b]{0.8\textwidth}
			\includegraphics[width=120mm,height=24mm]{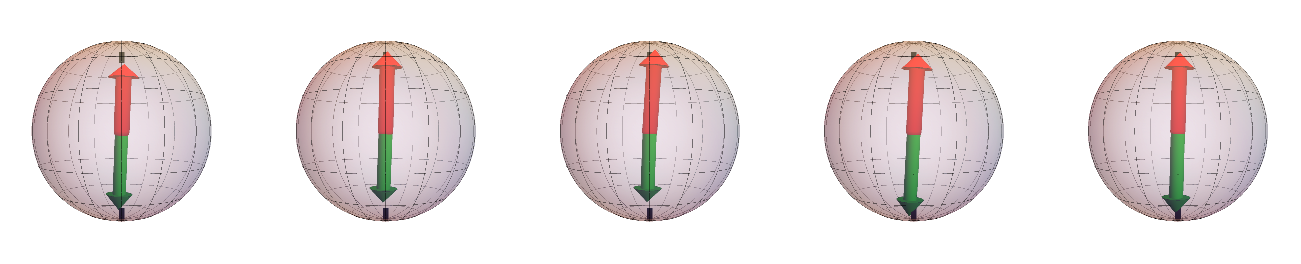}
			\caption{Parallel measurement on IBM Q 5 Yorktown.}
		\end{subfigure}
		\captionsetup{justification=raggedright}
		\caption{Detector spheres for qubits 0 to 4 of (a), (b) IBM Q 5 Tenerife (c), (d) IBM Q 5 Yorktown. The arrow represents the vector $(a_1,a_2,a_3)/a_0$ from measurment $\tilde{\Pi}^{(n=0,1)}=\vec{a}^{(n)}\cdot \vec{\sigma}$, where the north pole and the south pole correspond to ideal $|0\rangle\langle0|$ and $|1\rangle\langle 1|$ respectively. Positivity is reflected by the length of the arrow being smaller than 1. The width of the arrow represents the weight $a_0$ in the corresponding POVM element, for which the ideal case is $1/2$.} 
		\label{fig:DetectorSpheres}
	\end{figure*}
	
	\begin{figure*}
		\centering
		\begin{subfigure}[b]{1.0\textwidth}
			\includegraphics [width=\textwidth]{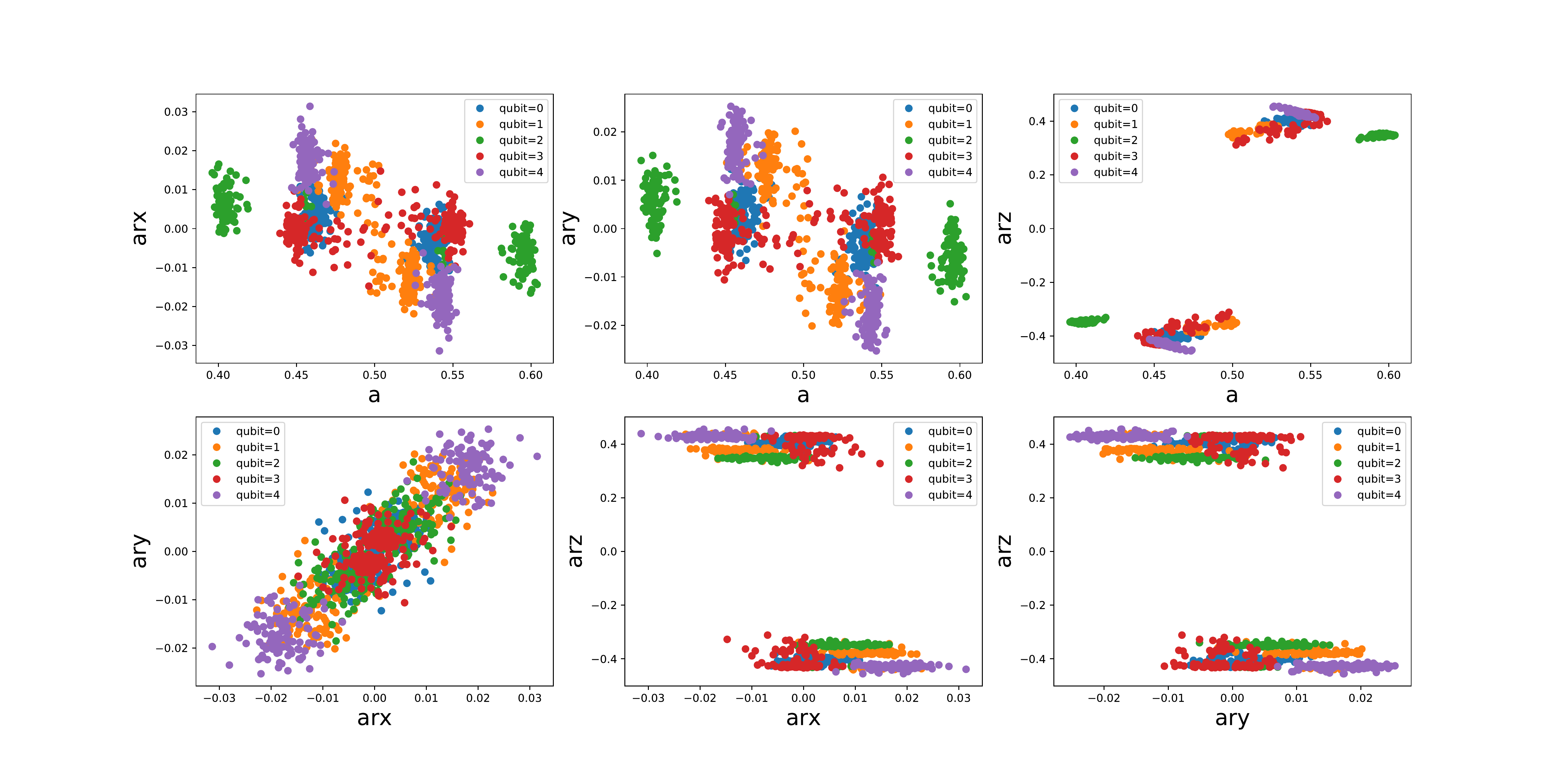}
			\caption{Individual measurement.}
		\end{subfigure} \\
		\centering
		\begin{subfigure}[b]{1.0\textwidth}
			\includegraphics [width=\textwidth]{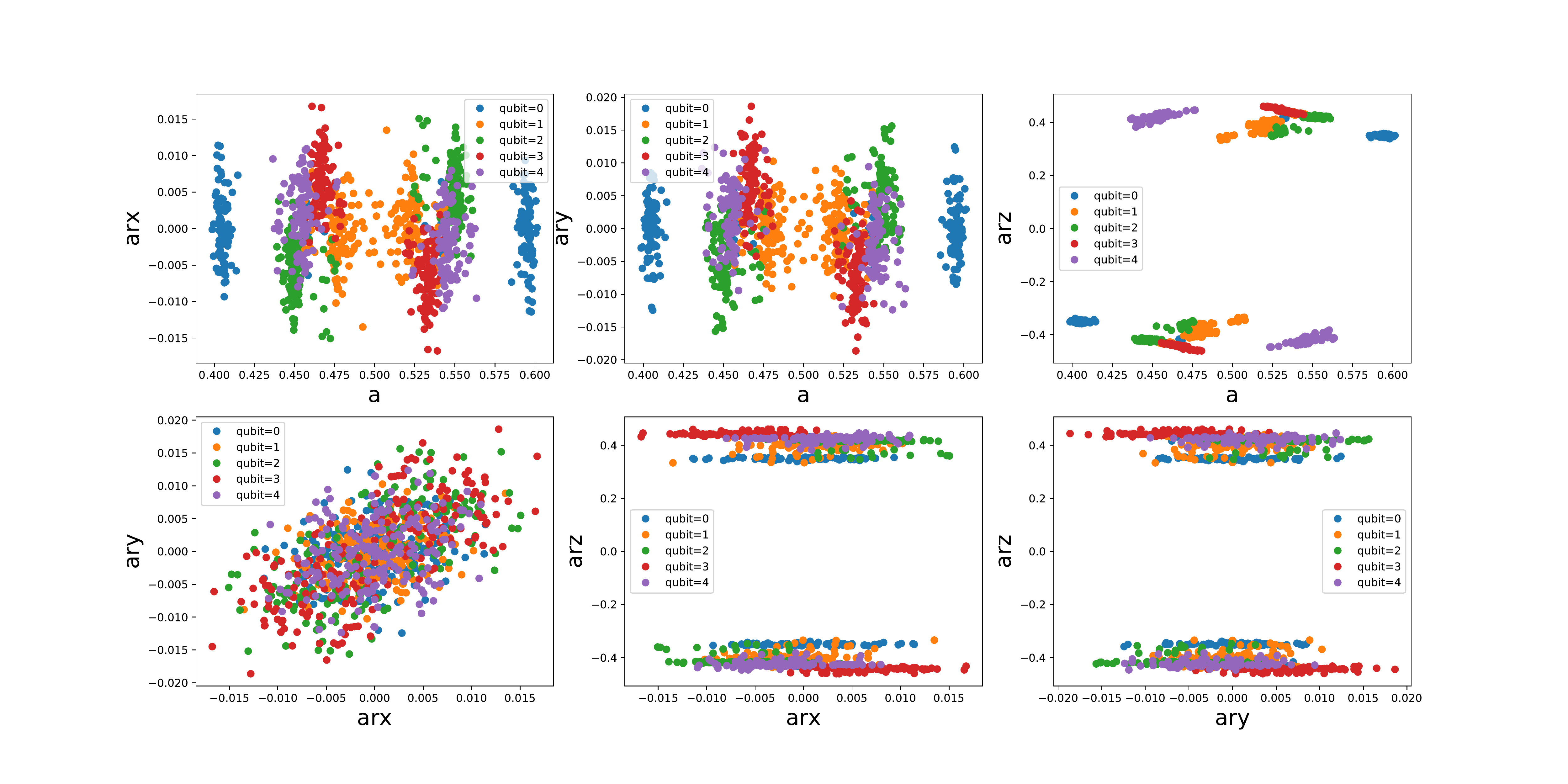}
			\caption{Parallel measurement.}
		\end{subfigure}
		\captionsetup{justification=raggedright}
		\caption{Visualizing the single-qubit detector parameters for the five qubits in IBM Q 5 Tenerife obtained by (a) individual measurement and (b) parallel measurement. The parameters $ar_{x, y, z} = a_{1, 2, 3}$. These are displayed for both $\Pi^{(1)}$ and $\Pi^{(0)}$.}
		\label{fig:data_visual4}
	\end{figure*}

	We performed QDT on the two IBM Q 5  devices: Tenerife (ibmqx4) and  Yorktown (ibmqx2), and present the results below. The test states in ${\cal A}_1$ were prepared by first initializing the qubits in $\ket0$ (which is the ground state of each qubit) and acting on it by the single-qubit gates: Pauli $X$, Hadamard  $H$, and the $S$ gates, as well as their combinations. The MLE~\cite{Fiurasek2001} was used for calculating the POVM parameters from measured frequencies, reviewed earlier. The positivity is ensured by construction, e.g., using initial POVM elements being $\openone/2$ for the iteration. First we adopted the single-qubit detector model. One can carry out the detector tomography procedure for each physical qubit individually, leaving the other qubits in the machine idle, or simultaneously carry out the same procedure for all qubits (or a subset of them). We henceforth refer to these two different ways as `individual measurement' and `parallel measurement', respectively. In principle there should not be any difference except that due to statistical fluctuations between the two, since using the single-qubit detector model we have assumed independence of the qubits. However, in reality we see significant discrepancy between the results obtained from the two types of experiments, which we will describe below. 
	
	A next-step generalization would be to adopt the two-qubit detector model. We examined all pairs of qubits in the two machines, and compared the results with those obtained for single-qubit detector model. If the discrepancy we observed is solely due to pairwise influence, this would be captured in the two-qubit detector tomography. However, this is not the case, as we will see in Sec.~\ref{sec:double}. One can readily generalize this to detector models involving three or more qubits as in Eq.~\ref{eq:multiqubit}. For the five-qubit devices, a five-qubit detector model will be the best to characterize the measurement for the two 5-qubit IBM machines. In order to obtain all $2^5=32$ operators $\Pi^{(\vec{n})}$ using the aforementioned basis states, $6^5=7776$ circuits are required. But some kind of compressed sensing technique may be used to mitigate this, as was done for QST~\cite{GrossLiuFlammiaEtAl10}. We would like to point out that to run this list of circuits on the current devices, it needs to be separated into smaller lists of jobs, since there is an upper limit on the circuit count for one single submitted job.
	
	\subsection{Single qubit detector: parallel vs. individual}
		The results of QDT are visualized in Fig.~\ref{fig:DetectorSpheres} using Bloch spheres. The 3d arrow represents the vector $\vec{r}=(a_1,a_2,a_3)/a_0$, and should be $(0,0,\pm1)$ for ideal detectors $\Pi^{0/1}=(\openone \pm \sigma_z)/2$. We use the thickness of the arrow to represent the parameter $a_0$. 
	Each detector is found to have its axis align mostly with $z$ axis but behave with some notable difference to the ideal 0/1 projectors: (1) $a_0^{(n)}$ deviating from 1/2; (2) $a_3^{(n)}/a_0^{(n)}$ deviating from $\pm1$; (3) $a_{1,2}^{(n)}$ being slightly non-zero. These features are displayed in Fig.~\ref{fig:DetectorSpheres}. 
	
	In Fig.~\ref{fig:data_visual4}, we show scattered plots for detector parameters in IBM Q 5 Tenerife for 100 different runs. Results for the other device are in Fig.~\ref{fig:data_visual2}. We summarize the single-qubit detector results for the two devices, as measured individually for each physical qubit leaving the other qubits idle, in Table~\ref{tab:ibmqx4_ind} and Table~\ref{tab:ibmqx2_ind} respectively. And those obtained by carrying out single-qubit detector tomography for all five qubits in the machine simultaneously are presented in Table~\ref{tab:ibmqx4_para} and Table~\ref{tab:ibmqx2_para}. The error estimated for each parameter is typically of the order $O(10^{-4})$, with the largest error among them up to 0.003. Fig.~\ref{fig:data_visual4} corresponds to data in Table~\ref{tab:ibmqx4}, and Fig.~\ref{fig:data_visual2} corresponds to data in Table~\ref{tab:ibmqx2}.
		
	\begin{table}[H]
	\centering
	\begin{tabular}{c | c | c | c | c | c}
		\diagbox{device}{qubit} & 0 & 1 & 2 & 3 & 4 \\
		\hline
		ibmqx4 & 0.011 & 0.010 & 0.023 & 0.087  & 0.025 \\
		ibmqx2 & 0.042 & 0.017 & 0.044 & 0.031 & 0.024
	\end{tabular}
	\caption{Distance between single-qubit detector from individual measurement and that from parallel measurement, for IBM Q 5 Tenerife (ibmqx4) and IBM Q 5 Yorktown (ibmqx2).} \label{tab:ibmqx42_diff}
	\end{table} \vspace{5mm}
	A notable feature is found that for almost all qubits $a_0^{(0)}$ is larger than $a_0^{(1)}=1-a_0^{(0)}$, which comes from relaxation to the ground state $|0\rangle$. There is an exception for qubit 3 of the device IBM Q 5 Tenerife, where $a_0^{(0)} < a_0^{(1)}$ when measured together with the other qubits in parallel. This was not seen when qubit 3 was measured alone, which hints at influence from the other qubits.
	
	A measure of discrepancy between individual measurement and parallel measurement is the distance between the vectors $\vec{a}^{(0)}=(a_0^{(0)}, a_1^{(0)}, a_2^{(0)}, a_3^{(0)})$ (note that $\vec{a}^{(1)}=(1,0,0,0)-\vec{a}^{(0)}$) obtained in the two different ways. This distance corresponds to the Frobenius norm of the difference between the two $\Pi^{(0)}$ operators up to a factor of 2. These are shown in Table~\ref{tab:ibmqx42_diff}. It is worth noticing that the distance associated with statistical fluctuations in the estimated $\vec{a}^{(0)}$ is typically of the order $O(10^{-3})$. We see that the distance between the two $\vec{a}^{(0)}$ vectors obtained from individual measurement and parallel measurement is one order of magnitude larger, which indicates that there is some correlation due to several qubits being operated and measured simultaneously, visible even in the presence of statistical fluctuations.
		
	\subsection{Beyond single qubit detector} \label{sec:double}
		
	\begin{table}
	\centering
	\begin{subtable}[t]{0.5\textwidth}
	\begin{tabular}{c | c | c | c | c | c}
		\diagbox{ \begin{tabular}{l}qubit of \\ interest\end{tabular}}{
		\begin{tabular}{l} qubit \\traced out\end{tabular}} & 0 & 1 & 2 & 3 & 4 \\
		\hline
		0 & - & 0.0006 & 0.0006 & 0.0004 & 0.0003 \\
		\hline
		1 & 0.0004 & - & 0.0154 & 0.0005 & 0.0001 \\
		\hline
		2 & 0.0008 & 0.0016 & - & 0.0009 & 0.0023 \\
		\hline
		3 & 0.0002 & 0.0006 & 0.1642 & - & 0.0028 \\
		\hline
		4 & 0.0010 & 0.0007 & 0.0022 & 0.0023 & - \\
	\end{tabular}
	\caption{IBM Q 5 Tenerife.}
	\end{subtable} \\
	\vspace{5mm}
	\begin{subtable}[t]{0.5\textwidth}
	\begin{tabular}{c | c | c | c | c | c}
		\diagbox{ \begin{tabular}{l}qubit of \\ interest\end{tabular}}{
		\begin{tabular}{l} qubit \\traced out\end{tabular}} & 0 & 1 & 2 & 3 & 4 \\
		\hline
		0 & - & 0.0013 & 0.0036 & 0.0007 & 0.0010 \\
		\hline
		1 & 0.0011 & - & 0.0116 & 0.0009 & 0.0133 \\
		\hline
		2 & 0.0033 & 0.0055 & - & 0.0090 & 0.0010 \\
		\hline
		3 & 0.0003 & 0.0006 & 0.0044 & - & 0.0191 \\
		\hline
		4 & 0.0007 & 0.0088 & 0.0020 & 0.0130 & - \\
	\end{tabular}
	\caption{IBM Q 5 Yorktown.}
	\end{subtable}
	\caption{Distance between single-qubit detector from pairwise parallel measurement and that obtained by tracing out another qubit in a two-qubit detector, for the two devices. The entry in $i$-th row and $j$-th column is the distance between single-qubit detector of qubit $i$ conditioned on qubit $j$ and that obtained from pairwise parallel measurements of the pair $i, j$.} \label{tab:ibmqx42_comp_pp}
	\end{table}
	
	\begin{table}
	\centering
	\begin{subtable}[t]{0.5\textwidth}
	\begin{tabular}{c | c | c | c | c | c}
		\diagbox{ \begin{tabular}{l}qubit of \\ interest\end{tabular}}{
		\begin{tabular}{l} qubit \\traced out\end{tabular}} & 0 & 1 & 2 & 3 & 4 \\
		\hline
		0 & - & 0.086 & 0.088 & 0.086 & 0.086 \\
		\hline
		1 & 0.030 & - & 0.040 & 0.021 & 0.012 \\
		\hline
		2 & 0.026 & 0.026 & - & 0.013 & 0.009 \\
		\hline
		3 & 0.028 & 0.016 & 0.195 & - & 0.012 \\
		\hline
		4 & 0.060 & 0.035 & 0.012 & 0.018 & - \\
	\end{tabular}
	\caption{Individual measurement.} \label{tab:ibmqx4_comp_ind}
	\end{subtable} \\
	\vspace{5mm}
	\begin{subtable}[t]{0.5\textwidth}
	\begin{tabular}{c | c | c | c | c | c}
		\diagbox{\begin{tabular}{l}qubit of \\ interest\end{tabular}}{\begin{tabular}{l} qubit \\traced out\end{tabular}} & 0 & 1 & 2 & 3 & 4 \\
		\hline
		0 & - & 0.079 & 0.082 & 0.080 & 0.080 \\
		\hline
		1 & 0.025 & - & 0.044 & 0.018 & 0.008 \\
		\hline
		2 & 0.010 & 0.018 & - & 0.013 & 0.015 \\
		\hline
		3 & 0.083 & 0.083 & 0.149 & - & 0.081 \\
		\hline
		4 & 0.046 & 0.029 & 0.016 & 0.007 & - \\
	\end{tabular}
	\caption{Parallel measurement.} \label{tab:ibmqx4_comp_par}
	\end{subtable}
	\caption{Distance between single-qubit detector from (a) individual measurement / (b) parallel measurement and that obtained by tracing out another qubit in a two-qubit detector, for IBM Q 5 Tenerife. The entry in $i$-th row and $j$-th column is the distance between single-qubit detector of qubit $i$ conditioned on qubit $j$ and that for qubit $i$ obtained from parallel measurements of all qubits.} \label{tab:ibmqx4_comp}
	\end{table}
	
	\begin{table}
	\centering
	\begin{subtable}[t]{0.5\textwidth}
	\begin{tabular}{c | c | c | c | c | c}
		\diagbox{\begin{tabular}{l}qubit of \\ interest\end{tabular}}{\begin{tabular}{l} qubit \\traced out\end{tabular}} & 0 & 1 & 2 & 3 & 4 \\
		\hline
		0 & - & 0.032 & 0.044 & 0.028 & 0.032 \\
		\hline
		1 & 0.020 & - & 0.020 & 0.022 & 0.026 \\
		\hline
		2 & 0.017 & 0.011 & - & 0.019 & 0.008 \\
		\hline
		3 & 0.003 & 0.018 & 0.014 & - & 0.080 \\
		\hline
		4 & 0.004 & 0.026 & 0.024 & 0.027 & - \\
	\end{tabular}
	\caption{Individual measurement.} \label{tab:ibmqx2_comp_ind}
	\end{subtable} \\
	\vspace{5mm}
	\begin{subtable}[t]{0.5\textwidth}
	\begin{tabular}{c | c | c | c | c | c}
		\diagbox{\begin{tabular}{l}qubit of \\ interest\end{tabular}}{\begin{tabular}{l} qubit \\traced out\end{tabular}} & 0 & 1 & 2 & 3 & 4 \\
		\hline
		0 & - & 0.014 & 0.034 & 0.034 & 0.037 \\
		\hline
		1 & 0.009 & - & 0.013 & 0.018 & 0.022 \\
		\hline
		2 & 0.029 & 0.039 & - & 0.029 & 0.037 \\
		\hline
		3 & 0.034 & 0.040 & 0.018 & - & 0.053 \\
		\hline
		4 & 0.024 & 0.018 & 0.010 & 0.008 & - \\
	\end{tabular}
	\caption{Parallel measurement.} \label{tab:ibmqx2_comp_par}
	\end{subtable}
	\caption{Distance between single-qubit detector from (a) individual measurement / (b) parallel measurement and that obtained by tracing out another qubit in a two-qubit detector, for IBM Q 5 Yorktown. The entry in $i$-th row and $j$-th column is the distance between single-qubit detector of qubit $i$ conditioned on qubit $j$ and that for qubit $i$ obtained from parallel measurements of all qubits.} \label{tab:ibmqx2_comp}
	\end{table}
	
	\noindent {\bf Two-qubit QDT and cross talk}. The two-qubit detector model $\Pi_2^{(n_0,n_1)}$ for a pair of qubits is characterized by $64$ parameters, which can be organized into four $4\times4$ matrices $c_{i,j}^{(n_0,n_1)}$ for the four outcomes $(n_0,n_1) = (00),(01),(10),(11)$ respectively. Imagine we have two uncorrelated systems $A$ and $B$, where the POVM for the composition is $\{\Pi_{AB}^{(n_A,n_B)}=\Pi_A^{(n_A)}\otimes\Pi_B^{(n_B)}\}$. $\sum_{n_A}\Pi_A^{(n_A)}=\iden_A$ and the same condition for $B$ are satisfied independently. When we have no access to system $B$, we need to sum over all possible outcomes for $B$ to get $\{ \Pi_{A}^{(n_A)}\otimes\iden_B=\sum_{n_B}\Pi_{AB}^{(n_A,n_B)} \}$. We can then take the partial trace over $B$ to recover
	\beq
		\Pi_{A}^{(n_A)}=\frac{1}{\mathrm{dim}(B)} \mathrm{Tr}_B \sum_{n_B}\Pi_{AB}^{(n_A,n_B)},
	\eeq
	where $\mathrm{dim}(B)$ is the dimension of the Hilbert space for $B$. In doing so we are assuming that any state of $B$ is equally likely to occur, i.e., no information about $B$ is accessible. To check whether a pair of qubits are separable, we can calculate the singular values of the matrices $c_{i,j}^{(n_0,n_1)}$ for them. If the single-qubit detector assumption holds well, the operators can be decomposed in the following way:
	\beq
		\Pi_2^{(n_0,n_1)} = \Pi_1^{(n_0)} \otimes \Pi_1^{(n_1)},
	\eeq
	where $\Pi_1^{(n)}$ is a single-qubit detector operator. In this case there will be only one nonzero singular value for any of the four $c_{i,j}^{(n_0,n_1)}$ matrices. This is a direct analogy to characterization of the entanglement of a bipartite system. We can also calculate from the singular values the analogy of entanglement measures, whose magnitudes give a measure of how bad the assumption of independent single-detectors is violated. We will not present detailed analysis about this here.

	How do we characterize a single-qubit detector reduced from the detector in the presence of other qubits? From the two-qubit detector model (see Eq.~\ref{eq:multiqubit} applied to two qubits), one can trace out one qubit and obtain a single-qubit detector model for the other qubit. For example, tracing out the second qubit in a pair, we get a single-qubit detector for the first qubit according to
	\beq
		\Pi_1^{(0)} = \frac{1}{2} \mathrm{Tr_{2nd}} \big( \Pi_2^{(00)}+\Pi_2^{(01)} \big),
	\eeq
	where the trace is taken over the second qubit only. Note that by doing this, $\Pi_1^{(0)}+\Pi_1^{(1)}=\iden_{\mathds{C}^2}$ is an automatic consequence of $\sum_{n_0,n_1} \Pi_2^{(n_0,n_1)}=\iden_{\mathds{C}^2\times\mathds{C}^2}$. We call the single-qubit detector obtained by tracing out another qubit in a pair `a single-qubit detector conditioned on another qubit'. We calculate such models for each qubit conditioned on any of the other qubits, and compare the result to the single-qubit detector obtained from parallel measurement of only those two qubits (henceforth referred to as pairwise parallel measurement). This is listed in Table~\ref{tab:ibmqx42_comp_pp}. We can see that by tracing out one qubit in a pair, the double-qubit detector result is reduced to the single-qubit result obtained from pairwise parallel measurement within statistical fluctuations for most pairs. In Tenerife, one significant exception is qubit 3's detector under the influence of qubit 2's detector, and in Yorktown, qubit 1 detection is influenced by that of qubit 2. 

	We also compare the single-qubit detector conditioned on another qubit to the single-qubit result obtained from individual measurement and to parallel measurement. In the ideal case where all detectors are independent of each other, these three results should agree within statistical uncertainty. If there is influence of only one other qubit on a given qubit, then we expect one of the four conditional single-qubit detector results to coincide with the result obtained from parallel measurement. From this we can also find which qubit is affecting a given qubit. Again we use the distance between two $\vec{a}^{(0)}$ vectors to characterize the agreement between two results. We organize the comparison into the tables.~\ref{tab:ibmqx4_comp} and \ref{tab:ibmqx2_comp}. As mentioned before, due to statistical fluctuations the distance can not be resolved below the order $O(10^{-3})$. Therefore, it is sensible to compare the entries in the tables to this order of magnitude. For all qubits in both Yorktown and Tenerfie, the distances obtained are at least one order of magnitude larger, and with some even larger. This suggests that pairwise influence and crosstalk do exist. This is important to take into account when we analyze results of measurement, and this suggests that by adopting the two-qubit detector model, the measurement result may be further improved than using just the single-qubit detector model.
	
	In Table~\ref{tab:ibmqx4_comp_ind} we can see that the result for qubit $3$ measured individually differs significantly from that conditioned on qubit $2$. This suggests possible influence on qubit $3$ by qubit $2$. If qubit $2$ is the single source of nontrivial effect, one would expect that the result for qubit $3$ measured in parallel with the others to agree with the result conditioned on qubit $2$ (operating only qubits 2 and 3). From Table~\ref{tab:ibmqx4_comp_par}, however, we can see that this is not the case. In fact, the result for qubit $3$ measured in parallel with all the other qubits differs from any of the results obtained by tracing out the other qubit from two-qubit detectors. This implies nontrivial correlation when several physical qubits are being operated on. 
	
	\medskip\noindent {\bf Three-qubit QDT}. 	As discussed in Sec.~\ref{sec:main}, this procedure can be generalized to more qubits. Ideally one would use a five-qubit detector model in the two IBM Q 5 devices, but for the purpose of demonstration, we used a triple-qubit detector model to characterize three physical qubits that are `connected' in the sense that a CNOT gate can be applied to any pair among them. The result for a triplet of qubits is characterized by $512$ parameters, which can be organized into eight $4\times4\times4$ matrices for the four outcomes $000,001,010,011,100,101,110,111$ respectively. Due to the considerably larger number of circuits required, we only repeat these experiments for 50 runs (each with 8192 shots). We used these data to demonstrate the first-step correction of experimental data in Sec.~\ref{sec:correction}.

	\subsection{Error analysis}
	
	Following the non-parametric bootstrap error analysis adopted in \cite{Blume-Kohut2017}, we evaluated the uncertainty in the parameters by first obtaining different estimates from resampled data sets of experimental data and calculating statistics of these estimates. In our analysis each experiment was repeated 100 times, each time with 8192 shots on the IBM Q devices. This gives us effectively 819200 shots, from which the result is calculated using MLE. To evaluate an uncertainty in this result, we resample the set of 100 runs with replacement to obtain new sets of experimental data. These sets are of the same size (100) as our original data set and from each we can calculate a new estimate of the result. The standard deviation of these values gives an estimate of the uncertainty in our result due to statistical fluctuation. 100 resampled data sets were generated for each experiment. For some selected cases we tried more (up to 1000) resampled data sets, which gave similar standard deviation to that from 100 resampled data sets. Therefore, we believe the standard deviation from 100 resampled data sets can represent the fluctuations well.
	
	\section{Application of characterized detectors---Inferring ideal detection}
	\label{sec:ideal}
	Given the characterized detectors, one should be able to infer from the existing measurement data the `correct' joint distribution $P_{(n_0, n_1, ..., n_{N-1})}$ of obtaining $N$-qubit outcomes $(n_0, n_1, ..., n_{N-1})$ in the ideal computational basis to some extent. Assuming there is no detector crosstalk,
	\begin{eqnarray}
		\tilde{P}_{(n_0, n_1, ..., n_{N-1})} &=& {\rm Tr}(\rho\,\tilde{ \Pi}_{[0]}^{n_0}\otimes \tilde{\Pi}_{[1]}^{n_1}\otimes ...\otimes \tilde{\Pi}_{[N-1]}^{n_{N-1}}) \nonumber \\
			&=& {\rm Tr} \big[ \rho\,\prod_{j=0}^{N-1} \big( \sum_{q=0}^3 a_{q,[j]}^{(n_j)} \sigma_{q,[j]} \big) \big],
		\label{eq:pMp}
	\end{eqnarray}
	where $[j]$ denotes the $j$-th physical qubit in the device; we use $\tilde{P}_{(n_0, n_1, ..., n_{N-1})}$ to denote the experimental distribution and $P_{(n_0, n_1, ..., n_{N-1})}$ the ideal distribution. And when $|a_1^{(n)}|, |a_2^{(n)}| \ll |a_3^{(n)}|$, which is the case in the IBM's quantum computers, Eq.~\ref{eq:pMp} becomes approximately:
	\begin{widetext}
	\begin{eqnarray}
		\tilde{P}_{(n_0, n_1, ..., n_{N-1})} &\approx& {\rm Tr} \big[ \rho\,\prod_{j=0}^{N-1} (a_{0,[j]} + a_{3,[j]}^{(n_j)}\sigma_{3,[j]}) \big] \nonumber \\
			&\approx& \sum_{(m_0, m_1, ..., m_{N-1})} P_{(m_0, m_1, ..., m_{N-1})} \prod_{j=0}^{N-1} \Big(a_{0,[j]}^{(n_j)} + (-1)^{m_j}a_{3,[j]}^{(n_j)}\Big), \nonumber \\
			&\approx& \sum_{\vec{m}}M_{\vec{n}; \vec{m}} P_{\vec{m}},
		\label{eq:approx_pMp}
	\end{eqnarray}
	\end{widetext}
	where summation over $\vec{m}$ runs through all possible outcomes. Moreover, $M$ is a left-stochastic matrix and its matrix elements are given by
	\beq
		M_{\vec{n}; \vec{m}} \equiv \prod_{j=0}^{N-1} \Big(a_{0,[j]}^{(n_j)} + (-1)^{m_j}a_{3,[j]}^{(n_j)}\Big).
	\eeq	
	This can be used to invert the relation to obtain $P_{\vec{m}}$. 
	
	However, a problem is that the resultant $P_{\vec{m}}$ by direct inversion may have negative components. Similar issues were also addressed in~\cite{Blume-Kohut2013}, and dealt with by setting a cutoff. In the near-term devices (e.g. IBM Q 5 Yorktown in Sec.~\ref{sec:correction}), this problem is very likely to occur due to statistical fluctuations in measured frequencies. Another way to obtain $P_{\vec{m}}$ circumventing the negativity problem is to minimize the distance squared $|MP-\tilde{P}|^2$, subject to the constraints of positivity and normalization. This is a quadratic programming problem, whose objective function is convex (and solution can be found in polynomial time using ellipsoid method). In Sec.~\ref{sec:correction} we demonstrate this procedure using a built-in function in the python package Scipy~\cite{Scipy}. We stress that this correction procedure only serves as an easy first-step mitigation, which does not have the full power of QST using characterized detectors. The advantage is that one only needs the measured frequencies for all outcomes in the computational basis and no further experiments are needed.
	
	We remark that this conclusion is based on the assumption that $\tilde{\Pi}^{(n_j)}_{[j]}\approx a_{0,[j]}^{(n_j)} \iden + a_{3,[j]}^{(n_j)} \sigma_{3,[j]}$. The situation will be complicated when there are non-negligible components $a_1$ and $a_2$, in which case a trick can be used if we can run additional circuits, which are the same as before except with additional Pauli $Z$ gates at the end. This is similar to the idea behind the error mitigation scheme in~\cite{Calderon-Vargas2017}. The gates added to the end are of the form:
	\beq
	\label{eqn:ZK}
		Z(\vec{K}) \equiv \prod_{i=0}^{N-1} \sigma_{3,[i]}^{K_i},
	\eeq
	where $\vec{K}$ is a binary string of length $N$ that denotes whether there is a Pauli $Z$ gate on each qubit in the device. Given a particular $\vec{K}$, the probability is given by:
	\begin{widetext}
	\begin{eqnarray}
		\tilde{P}_{\vec{n}}(\vec{K}) &=& {\rm Tr} \big[ Z(\vec{K}) \rho Z(\vec{K})\,\prod_{j=0}^{N-1} \big( \sum_{q=0}^3 a_{q,[j]}^{(n_j)} \sigma_{q,[j]} \big) \big] \nonumber \\
			&=& {\rm Tr} \big[ \rho\,  \prod_{j=0}^{N-1} \big( (a_{0,[j]}^{(n_j)} + a_{3,[j]}^{(n_j)}\sigma_{3,[j]}) + (-1)^{K_j} (a_{1,[j]}^{(n_j)}\sigma_{1,[j]} + a_{2,[j]}^{(n_j)}\sigma_{2,[j]}) \big) \big].
	\end{eqnarray}
	\end{widetext}
	There are $2^N$ different $\vec{K}$'s, including the original circuit with the probability given by Eq.~\ref{eq:pMp}. Adding up $\tilde{P}_{\vec{n}}(\vec{K})$'s cancels the terms involving $a_1, a_2$ and their average gives the probability in Eq.~\ref{eq:approx_pMp}.
	
	Crosstalk between qubits can further complicate the situation. Let us again make the assumption that in Eq.~\ref{eq:multiqubit} only coefficients involving Pauli indices $i=0, 3$ dominate, i.e., $c^{(\vec{n})}_{\vec{i}} \approx 0$ for all the $\vec{i}$'s with any entry equal to 1 or 2. Now the probability is:
	\begin{eqnarray}
		\tilde{P}_{\vec{n}} &=& \sum_{i_0=0, 3} ... \sum_{i_{N-1}=0, 3} c^{(\vec{n})}_{\vec{i}} {\rm Tr} \big( \rho \sigma_{i_0} \otimes ... \otimes \sigma_{i_{N-1}} \big) \nonumber \\
			&=& \sum_{i_0=0, 3} ... \sum_{i_{N-1}=0, 3} c^{(\vec{n})}_{\vec{i}} \sum_{\vec{m}} (-1)^{\vec{m} \cdot \vec{i}/3} P_{\vec{m}}, \nonumber \\
			&=& \sum_{\vec{m}} \hat{M}_{\vec{n}; \vec{m}} P_{\vec{m}},
		\label{eq:approx_pMp_N}
	\end{eqnarray}
	where $\hat{M}$ is given by:
	\begin{eqnarray}
		\hat{M}_{\vec{n}; \vec{m}} = \sum_{\vec{I}} c^{(\vec{n})}_{\vec{I}} (-1)^{\vec{m} \cdot \vec{I}/3}.
	\end{eqnarray}
	Although $\vec{I}=(i_0, ..., i_{N-1})$ has the same expression as $\vec{i}$, we distinguish between them because each component of $\vec{I}$  is equal to 0 or 3. Note the summation only runs through these $\vec{I}$'s. We can use the same procedure to extract the ideal distribution $P_{\vec{m}}$. Moreover, if the assumption that $c^{(\vec{n})}_{\vec{i}} \approx 0$ for all the $\vec{i}$'s with any entry equal to 1 or 2 does not hold, we can still use the average procedure by running additional circuits with gates (\ref{eqn:ZK}) appended to the original circuits.
		
	\subsection{Using characterized detectors} \label{sec:correction}

	We demonstrate how to apply the characterized detectors in a simple real-life experiment for a first-step correction, without carrying out QST, as described in Sec.~\ref{sec:main}. First we applied Hadamard gate on qubit 3, and then CNOT gate on qubits 3 (condition) and 4 (target) in IBM Q 5 Yorktown, followed by measurement of all qubits. The ideal probability distribution will be $P_{00000}=P_{11000}=0.5$ with all other components of $P_{\vec{n}}$ equal to 0. The circuit was repeated for 50 runs, each run with 8192 shots. The largest two components are $\tilde{P}_{00000}=0.466$ and $\tilde{P}_{11000}=0.422$, with the others ranging from 0 to the order 0.01 ($\tilde{P}_{00001}=0.013$, $\tilde{P}_{01000}=0.042$, $\tilde{P}_{10000}=0.032$, $\tilde{P}_{11001}=0.011$). Direct inversion gives some negative entries in $P_{\vec{n}}$. An immediate technique is setting any negative entry to zero, and then renormalizing $P_{\vec{n}}$. This results in the two largest components being $\tilde{P}_{00000}=0.479$ and $\tilde{P}_{11000}=0.498$, with the biggest among the others of the order 0.01. We then turn to maximizing $|MP-\tilde{P}|^2$ subject to constraints of positivity and normalization. We argue that this method is more desirable because it avoids setting some arbitrary small value as the cutoff. This was done using the optimization function `optimize.minimize' in the python package Scipy, with the sequential least squares programming (SLSQP) method. The parameter `ftol' was set to $10^{-20}$ and optimization was typically done after between 300 and 400 iterations. First we apply Eq.~\ref{eq:approx_pMp} with detector parameters obtained from individual measurement. The `corrected' $P_{\vec{n}}$ has two dominant components, $P_{00000}=0.493$ and $P_{11000}=0.507$, with all the other components of the order $O(10^{-17})$. We repeat this analysis using detector parameters obtained from parallel measurement, which gives $P_{00000}=0.495$ and $P_{11000}=0.505$ and the other components of the order $O(10^{-17})$.
	
	To demonstrate the use of the double-qubit and triple-qubit detector models, we prepared a Bell state on two qubits or a GHZ state on three qubits using Hadamard gate and CNOT gate, followed by measuring only the two or three qubits involved. It is worth noting that the CNOT gates used respect the connectivity of the qubits in the real machine, so that we know which qubits are actually operated on. We list the measured frequencies in comparison with their corrected versions in Table~\ref{tab:correction23}. It is clear that with the first-step correction $P_{00}$ and $P_{11}$ (or $P_{000}$ and $P_{111}$ in the triple-qubit cases) are brought closer to the ideal value $0.5$; and their difference is reduced. We also note that the direct inversion with cutoff at zero and the optimization using SLSQP give similar results.
	
	\begin{table*}
	\begin{subtable}[t]{1.0\textwidth}
	\begin{tabular}{c | c | c | c | c | c | c | c}
		qubits of interest & control-target & data type & $P_{00}$ & $P_{01}$ & $P_{10}$ & $P_{11}$ & $|MP-\tilde{P}|$ \\
		\hline
		0, 1 & 0-1 & experiment & 0.470 & 0.040 & 0.054 & 0.436 & - \\
		\hline
		0, 1 & 0-1 & inversion & 0.489 & 0.003 & 0.0 & 0.509 & 0.0055 \\
		\hline
		0, 1 & 0-1 & optimization & 0.490 & 0.001 & 0.0 & 0.509 & 0.0053 \\
		\hline
		3, 4 & 3-4 & experiment & 0.481 & 0.031 & 0.041 & 0.448 & - \\
		\hline
		3, 4 & 3-4 & inversion & 0.480 & 0.024 & 0.0 & 0.497 & 0.0158 \\
		\hline
		3, 4 & 3-4 & optimization & 0.483 & 0.019 & 0.0 & 0.498 & 0.0150
	\end{tabular}
	\caption{2-qubit Bell state $\frac{1}{\sqrt{2}}(\ket{00}+\ket{11})$, corrected using the characterized double-qubit models.} \label{tab:correction2}
	\end{subtable} \\
	\vspace{5mm}
	\begin{subtable}[t]{1.0\textwidth}
	\begin{tabular}{c | c | c | c | c | c | c | c | c | c | c | c}
		qubits of interest & control-target & data type & $P_{000}$ & $P_{001}$ & $P_{010}$ & $P_{011}$ & $P_{100}$ & $P_{101}$ & $P_{110}$ & $P_{111}$ & $|MP-\tilde{P}|$ \\
		\hline
		0, 1, 2 & 0-1, 1-2 & experiment & 0.466 & 0.033 & 0.008 & 0.043 & 0.006 & 0.014 & 0.068 & 0.362 & - \\
		\hline
		0, 1, 2 & 0-1, 1-2 & inversion & 0.485 & 0.014 & 0.000 & 0.035 & 0.004 & 0.000 & 0.026 & 0.437 & 0.0040 \\
		\hline
		0, 1, 2 & 0-1, 1-2 & optimization & 0.487 & 0.013 & 0.000 & 0.034 & 0.002 & 0.000 & 0.025 & 0.438 & 0.0032 \\
		\hline
		2, 3, 4 & 3-4, 4-2 & experiment & 0.476 & 0.007 & 0.016 & 0.018 & 0.012 & 0.052 & 0.050 & 0.370 & - \\
		\hline
		2, 3, 4 & 3-4, 4-2 & inversion & 0.482 & 0.003 & 0.015 & 0.013 & 0.000 & 0.046 & 0.029 & 0.412 & 0.0076 \\
		\hline
		2, 3, 4 & 3-4, 4-2 & optimization & 0.485 & 0.003 & 0.014 & 0.011 & 0.000 & 0.044 & 0.029 & 0.415 & 0.0059
	\end{tabular}
	\caption{3-qubit GHZ state $\frac{1}{\sqrt{2}}(\ket{000}+\ket{111})$, corrected using the characterized triple-qubit models.} \label{tab:correction3}
	\end{subtable}
	\caption{Measured frequencies in comparison with those after inversion with cutoff at zero, and with those after minimizing $|MP-\tilde{P}|^2$, for (a) a 2-qubit state and (b) a 3-qubit state prepared on IBM Q 5 Yorktown. The experiment was repeated for 50 runs, each run with 8192 shots. The leftmost column is the qubits operated on and measured.} \label{tab:correction23}
	\end{table*}
	
	\section{Gate set tomography}
\label{sec:GST}
	Here we provide results of the GST analysis we carry out on the two IBM machines. We begin with a brief review of the GST scheme developed in~\cite{Blume-Kohut2013, Blume-Kohut2017}. A gate set in GST is defined as the collection of an unknown initial state $\rho$, a set of unknown CPTP gates $\{G_k\}$, and a 2-outcome unknown POVM $\{E, \iden-E\}$. The first step is called `linear GST', which broadly speaking is to express the gate set in some arbitrary basis. Taking the Hilbert-Schmidt space of matrices on the original Hilbert space as the new vector space, the expressions are the following:
	\begin{eqnarray}
		& \tilde{\iden} = \sum_{j,k}\langle \bra EF_jF_k\ket\rho\rangle \nonumber \\
		& \hat{\ket\rho}\rangle = \tilde{\iden}^{-1} \sum_j \langle\bra EF_j\ket\rho\rangle \nonumber \\
		& \langle\hat{\bra E} = \sum_k \langle\bra EF_k\ket\rho \rangle\nonumber \\
		& \hat{G_i} = \tilde{\iden}^{-1} \sum_{j,k} \langle\bra EF_jG_iF_k\ket\rho\rangle,
	\end{eqnarray}
	$|\rho\rangle\rangle$ and $\langle\langle E|$ are vectorized version of $\rho$ and $E$, respectively, and $\{F_j\}$ is a set of gates  (acting on vectorized version of density matrices and POVM) with which the initial state and the measurement will be informationally complete respectively. The inner products in these equations are obtained from experiments. This can only determine the gate set up to a transformation:
	\begin{eqnarray}
		& \rho=M\hat{\rho} \nonumber \\
		& E=\hat{E}M^{-1} \nonumber \\
		& G_i=M\hat{G_i}M^{-1},
	\end{eqnarray}
	where $M$ is some invertible matrix. This freedom was termed `gauge' by the authors of~\cite{Blume-Kohut2013}, and can be removed by trying to match the gate set towards some target. Long sequences of gates are used in experiments so as to capture the amplified gate parameter errors and thus achieve better characterization of a certain subset of gates $\{G_k\}$, called `germs'. However,  SPAM parameter errors cannot be amplified. The full GST analysis in~\cite{Blume-Kohut2017} can be summarized as follows: 1) linear GST for the first estimate; 2) gauge optimization to match the target gate set; 3) iteratively adding data for $\chi^2$ minimization to avoid local minima; 4) final MLE analysis and gauge optimization.

	\subsection{Results of single-qubit detector characterization from GST}
	We also used GST and ran corresponding circuits on IBM Qx4, with the standard gate set ${\bf G}=\{G_x=R_x(\pi/2), G_y=R_y(\pi/2),G_I=\openone \}$ for the state preparation and measurement fiducials. GST is particularly well suited for characterizing gates but less so for state preparation and measurement. However, it still provides a good comparison for detector tomography without the bias of assuming high-fidelity gate and state preparation. To run the GST with long squence requires a large number of different circuits to run. Since we are only interested in the detectors, we only choose $\{G_I\}$ to be the germs and only of 6 different lengths, i.e., $[1,121,241,361,481,601]$.  (Our other motivation was to capture some simple relaxation or decoherence from the decohered identity operation; see also App.~\ref{sec:relaxation}.)
	Even for such a simple setup, there are 272 different circuits to run (compared to just 6 for QDT). For each we take 8192 shots to obtain statistics. Ideally, we could have included $G_x$ and $G_y$ in the set of germs, but it will require many more circuits to run.
	
We use the `pygsti' package (version 0.9.6) of python to analyze the data. We first run  linear-inversion GST (LGST) without using the long-sequence circuits to obtain an initial estimate. We perform the gauge optimization by setting gate parameters to be trace preserving (TP). If the obtained POVMs for the detectors are not positive, we repeat the analysis by additionally setting a nonzero SPAM penalty factor (typically from 0.3 to 0.5) so as to obtain positive POVMs.  We then project the gates (this does not modify characterization of the initial state nor measurement POVMs) to be completely positive TP (CPTP). Using such an initial estimate, we then run the long-sequence GST (LSGST), which takes into account long-sequence circuits by performing iterative maximum likelihood by including longer sequences successively. Since we are conerned mostly with the detecors' POVMs, if the obtained POVMs after LSGST are not positive, we will repeat gauge optimization by setting the spam penalty factor.  
	
The circuits for GST were done in parallel for all 5 qubits on both IBMQx2 and IBMQx4. The obtained detectors' POVMs characterized by the above GST procedure are shown in Tables~\ref{tab:ibmqx2_paraGST} and~\ref{tab:ibmqx4_paraGST}. They agree with a few percentages to those using simple detector tomography earlier. The particular qubit 3 of IBMQx4 also displays the unusual behavior that its detector $ a_0^{(n=0)}=0.4521$ is smaller than $a_0^{(n=1)}=0.5479$ in parallel measurement together with other qubits. 

We then performed individual GST procedure only for the qubit 3 (leaving all other qubits idle), and obtain the detectors' characterization $\vec{a}^{(n_3=0)}=(
	   0.5182, 0.0036, 0.0022, 0.4449) $ and
	   $\vec{a}^{(n_3=1)}
			 = (0.4818 ,  -0.0036 , -0.0022 , -0.4449)$, which is closer to what was obtained by simple detector tomography on the qubit 3 individually with about 4\% difference. 

\begin{table}[t!]
	\centering
	\begin{tabular}{c | c | c | c | c | c}
		qubit & operator & $a_0$ & $a_1$ & $a_2$ & $a_3$ \\
		\hline
		\multirow{2}{*}{0} & $\Pi^{(0)}$ & 0.5292 & -0.0116 & 0.0021 & 0.4707 \\
			& $\Pi^{(1)}$ & 0.4708 & 0.0116 & -0.0021 & -0.4707 \\
			\hline
		\multirow{2}{*}{1} & $\Pi^{(0)}$ & 0.5491 & 0.0046 & 0.0058 & 0.4594\\
			& $\Pi^{(1)}$ & 0.4509 &  -0.0046 & -0.0058 & -0.4594 \\
			\hline
		\multirow{2}{*}{2} & $\Pi^{(0)}$ & 0.5183 & 0.0013 & -0.0063 & 0.4816 \\
			& $\Pi^{(1)}$ & 0.4817 &-0.0013 & 0.0063 & -0.4816  \\
			\hline
		\multirow{2}{*}{3} & $\Pi^{(0)}$ & 0.4521 & 0.0064 & 0.0061 & 0.4520 \\
			& $\Pi^{(1)}$ & 0.5478 &  -0.0064 & -0.0061 & -0.4520 \\
			\hline
		\multirow{2}{*}{4} & $\Pi^{(0)}$ & 0.5006 & 0.0082 & 0.0079 & 0.4370 \\
			& $\Pi^{(1)}$ & 0.4994 &-0.0082 & -0.0079 & -0.4370 \\
	\end{tabular}
	\caption{Single-qubit detector results by using GST on IBM Q 5 Tenerife, measured for all five qubits in parallel. However, if we perform the same GST only on qubit 3, then we obtain  $\vec{a}^{(n_3=0)}=(
	   0.5182, 0.0036, 0.0022, 0.4449) $ and
	   $\vec{a}^{(n_3=1)}
			 = (0.4818 ,  -0.0036 , -0.0022 , -0.4449)$.
	 } \label{tab:ibmqx4_paraGST}
	\end{table}

	\begin{table}[t!]
	\centering
	\begin{tabular}{c | c | c | c | c | c}
		qubit & operator & $a_0$ & $a_1$ & $a_2$ & $a_3$ \\
		\hline
		\multirow{2}{*}{0} & $\Pi^{(0)}$ & 0.5074 & -0.0307 & -0.0298 & 0.4847 \\
			& $\Pi^{(1)}$ & 0.4926 & 0.0307 & 0.0298 & -0.4847 \\
			\hline
		\multirow{2}{*}{1} & $\Pi^{(0)}$ & 0.5107 & 0.0078 & 0.0061 & 0.4892 \\
			& $\Pi^{(1)}$ & 0.4893 & -0.0078 & -0.0061 & -0.4892   \\
			\hline
		\multirow{2}{*}{2} & $\Pi^{(0)}$ & 0.5131 & 0.0230 & 0.0229 & 0.4858 \\
			& $\Pi^{(1)}$ & 0.4869 & -0.0230 & -0.0229 & -0.4858   \\
			\hline
		\multirow{2}{*}{3} & $\Pi^{(0)}$ & 0.5137 & 0.0170 & 0.0136& 0.4858 \\
			& $\Pi^{(1)}$ & 0.4863 &   -0.0170 & -0.0136& -0.4858 \\
			\hline
		\multirow{2}{*}{4} & $\Pi^{(0)}$ &0.5206 &  0.0171 & 0.0139 & 0.4789 \\
			& $\Pi^{(1)}$ & 0.4794 & -0.0171 & -0.0139 & -0.4789  \\
	\end{tabular}
	\caption{Single-qubit detector results by using GST on IBM Q 5 Yorktown, measured for all five qubits in parallel.} \label{tab:ibmqx2_paraGST}
	\end{table}

	\section{Conclusion and discussion}
\label{sec:conclude}
	In summary we performed the standard quantum detector tomography on the two devices IBM Q 5 Tenerife and IBM Q 5 Yorktown, assuming negligible errors in the ground state preparation and the single-qubit gates used to prepare the eigenstates of the three Pauli operators. Our resultant POVM shows deviation from the ideal projectors $\{|0\rangle\langle0|, |1\rangle\langle1|\}$ and can be used for a first-order correction in experiments. We also found evidence of crosstalk between qubits in one device. In particular, discrepancy was seen between individual measurement and parallel measurement. We believe adopting two-, or more-qubit detector models can improve the results further compared with assuming independent single-qubit detector model.  To study that more knowledge about hardware is required. Some peculiar features were observed in the qubit 3 of IBM Q 5 Tenerife that need further investigation. 
	
	The peculiar behavior of the qubit 3 from simple QDT agrees with that obtained from using a more sophisticated approach of the GST. This method, in principle, is capable of deducing the initial state, gate operations, and measurement POVMs in one go, by running various circuit sequences. Our simple QDT relies on the assumption that the detector error rate is higher than that of state preparation and simple single-qubit gates. 
	
	We point out some directions for future work. Given that the total number $N$ of qubits can be large, complete detector tomography will not be efficient. One can consider employing compressed sensing, as done in the state tomography~\cite{GrossLiuFlammiaEtAl10}. On the other hand, since the characterization of the triad---detector, state and process---forms a loop, there should be further improvement (to the next order in error) on detector characterization, and then on the state and process tomography, an idea similar to that in~\cite{Keith2018}.

\begin{acknowledgments}
This work was supported by National Science Foundation under grant No. PHY  1620252 and a subcontract from Brookhaven National Laboratory. T.-C.W. also acknowledges support of a SUNY seed grant. We thank Anthony DeGennaro, Antonio Mezzacapo and  Ollie Saira for useful discussions. Y. C. would also like to thank Xinzhong Chen for discussion about error analysis.
\end{acknowledgments}
\bibliographystyle{unsrt}	
\bibliography{mybib}

\appendix

\section{Tables} \label{app:tables}

	\begin{figure*}
		\centering
		\begin{subfigure}[b]{1.0\textwidth}
			\includegraphics[width=\textwidth]{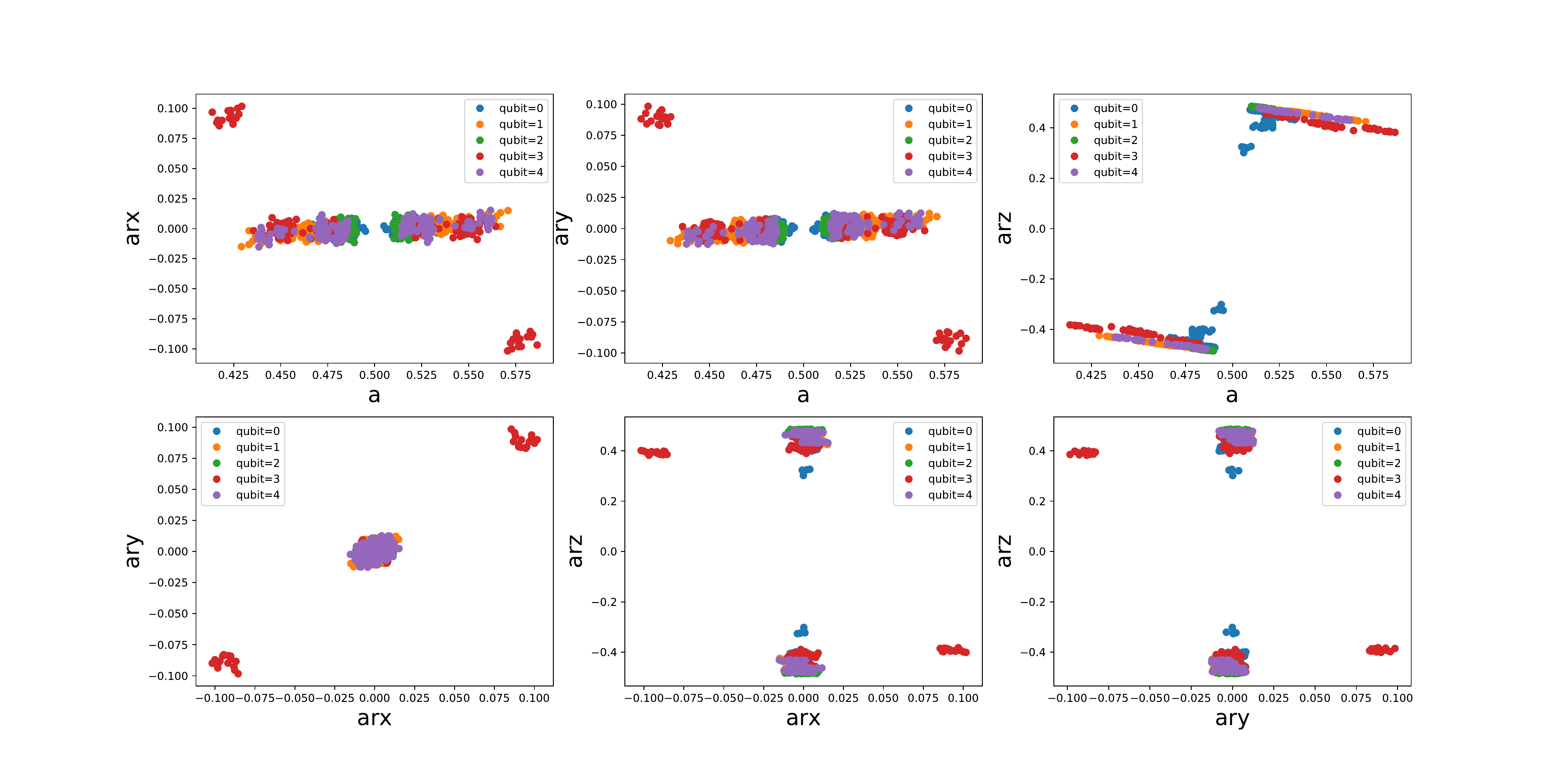}
			\caption{Individual measurement.}
		\end{subfigure} \\
		\begin{subfigure}[b]{1.0\textwidth}
			\includegraphics[width=\textwidth]{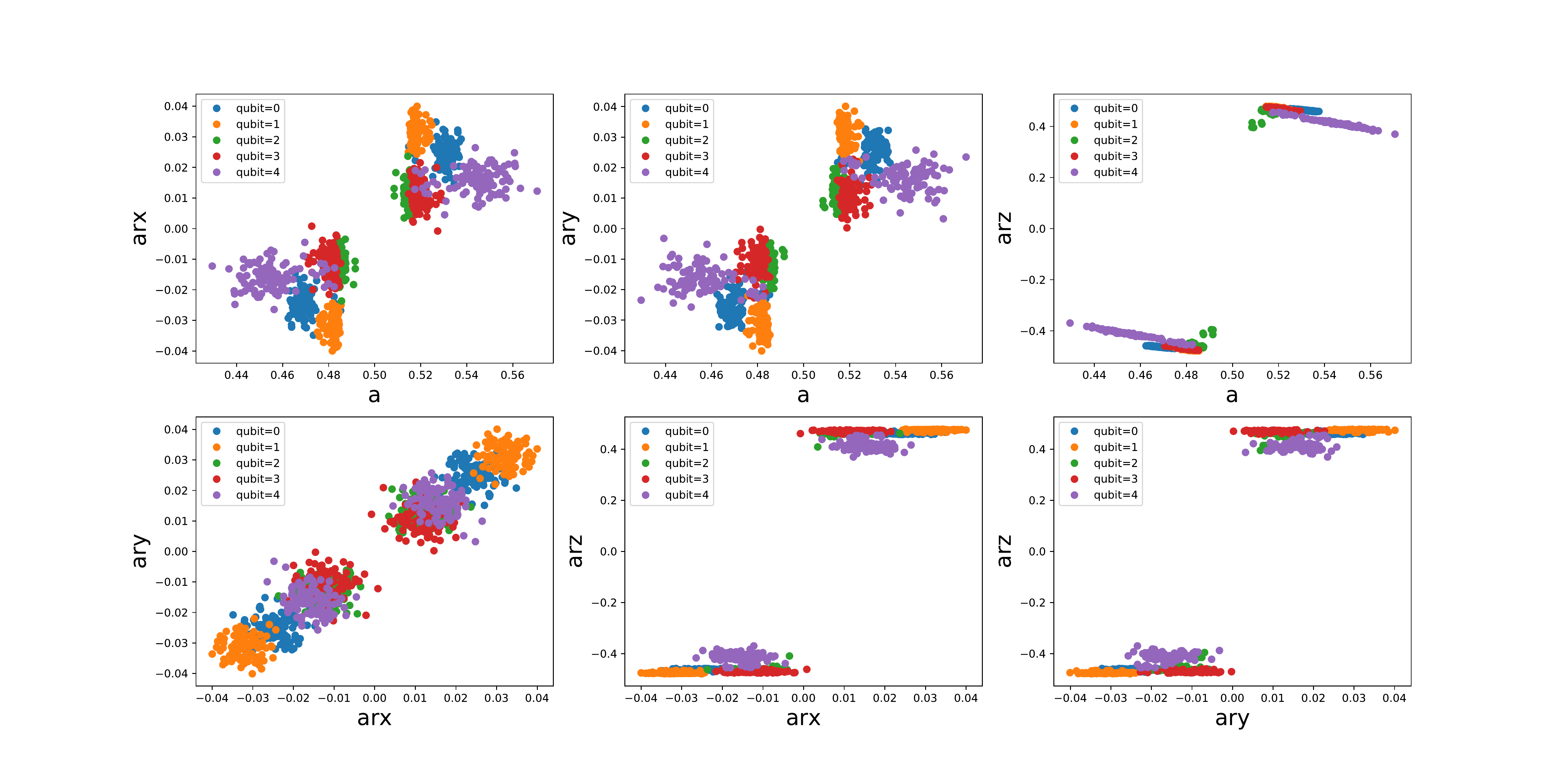}
			\caption{Parallel measurement.}
		\end{subfigure}
		\captionsetup{justification=raggedright}
		\caption{Visualizing the single-qubit detector parameters for the five qubits in IBM Q 5 Yorktown obtained by (a) individual measurement and (b) parallel measurement. The parameters $ar_{x, y, z} = a_{1, 2, 3}$. These are displayed for both $\Pi^{(1)}$ and $\Pi^{(0)}$.}
		\label{fig:data_visual2}
	\end{figure*}

	We list below the tables~\ref{tab:ibmqx4}, \ref{tab:ibmqx2} containing processed experimental data. The error estimated for each parameter is typically of the order $O(10^{-4})$, with the largest error among them up to 0.003. We adopt the precision according to the estimated errors. We also present the scattered plots for detector parameters in IBM Q 5 Yorktown for 100 different runs in Fig.~\ref{fig:data_visual2}, which correspond to data in Table~\ref{tab:ibmqx2}.

	\begin{table}
	\centering
	\begin{subtable}[t]{0.5\textwidth}
	\begin{tabular}{c | c | c | c | c | c}
		qubit & operator & $a_0$ & $a_1$ & $a_2$ & $a_3$ \\
		\hline
		\multirow{2}{*}{0} & $\Pi^{(0)}$ & 0.590(2) & -0.006(3) & -0.0063(4) & 0.3562(5) \\
			& $\Pi^{(1)}$ & 0.410(2) & 0.006(3) & 0.0063(4) & -0.3562(5) \\
			\hline
		\multirow{2}{*}{1} & $\Pi^{(0)}$ & 0.544(1) & 0.001(3) & 0.0008(3) & 0.4059(5) \\
			& $\Pi^{(1)}$ & 0.456(1) & -0.001(3) & -0.0008(3) & -0.4059(5) \\
			\hline
		\multirow{2}{*}{2} & $\Pi^{(0)}$ & 0.5427(5) & -0.0179(9) & -0.0173(5) & 0.4294(4) \\
			& $\Pi^{(1)}$ & 0.4573(5) & 0.0179(9) & 0.0173(5) & -0.4294(4) \\
			\hline
		\multirow{2}{*}{3} & $\Pi^{(0)}$ & 0.5381(5) & -0.003(1) & -0.0030(4) & 0.4054(4) \\
			& $\Pi^{(1)}$ & 0.4619(5) & 0.003(1) & 0.0030(4) & -0.4054(4) \\
			\hline
		\multirow{2}{*}{4} & $\Pi^{(0)}$ & 0.521(1) & -0.012(2) & -0.0122(4) & 0.3798(4) \\
			& $\Pi^{(1)}$ & 0.479(1) & 0.012(2) & 0.0122(4) & -0.3798(4)
	\end{tabular}
	\caption{Individual measurement.} \label{tab:ibmqx4_ind}
	\end{subtable} \\
	\vspace{5mm}
	\begin{subtable}[t]{0.5\textwidth}
	\begin{tabular}{c | c | c | c | c | c}
		qubit & operator & $a_0$ & $a_1$ & $a_2$ & $a_3$ \\
		\hline
		\multirow{2}{*}{0} & $\Pi^{(0)}$ & 0.587(2) & -0.000(3) & -0.0001(4) & 0.3618(5) \\
			& $\Pi^{(1)}$ & 0.413(2) & 0.000(3) & 0.0001(4) & -0.3618(5) \\
			\hline
		\multirow{2}{*}{1} & $\Pi^{(0)}$ & 0.5483(8) & 0.006(2) & 0.0053(4) & 0.4116(4) \\
			& $\Pi^{(1)}$ & 0.4517(8) & -0.006(2) & -0.0053(4) & -0.4116(4) \\
			\hline
		\multirow{2}{*}{2} & $\Pi^{(0)}$ & 0.5329(5) & -0.0065(7) & -0.0064(5) & 0.4430(5) \\
			& $\Pi^{(1)}$ & 0.4671(5) & 0.0065(7) & 0.0064(5) & -0.4430(5) \\
			\hline
		\multirow{2}{*}{3} & $\Pi^{(0)}$ & 0.4535(8) & 0.002(1) & 0.0023(4) & 0.4229(5) \\
			& $\Pi^{(1)}$ & 0.5465(8) & -0.002(1) & -0.0023(4) & -0.4229(5) \\
			\hline
		\multirow{2}{*}{4} & $\Pi^{(0)}$ & 0.522(1) & 0.000(2) & -0.0002(4) & 0.3975(4) \\
			& $\Pi^{(1)}$ & 0.478(1) & -0.000(2) & 0.0002(4) & -0.3975(4)
	\end{tabular}
	\caption{Parallel measurement.} \label{tab:ibmqx4_para}
	\end{subtable}
	\caption{Single-qubit detector results for IBM Q 5 Tenerife, (a) measured individually for each physical qubit leaving the other qubits idle; (b) measured for all five qubits in parallel.} \label{tab:ibmqx4}
	\end{table} \vspace{5mm}
	
	\begin{table}
	\centering
	\begin{subtable}[t]{0.5\textwidth}
	\begin{tabular}{c | c | c | c | c | c}
		qubit & operator & $a_0$ & $a_1$ & $a_2$ & $a_3$ \\
		\hline
		\multirow{2}{*}{0} & $\Pi^{(0)}$ & 0.545(2) & -0.013(2) & -0.012(3) & 0.424(3) \\
			& $\Pi^{(1)}$ & 0.455(2) & 0.013(2) & 0.012(3) & -0.424(3) \\
			\hline
		\multirow{2}{*}{1} & $\Pi^{(0)}$ & 0.530(2) & 0.003(1) & 0.0028(5) & 0.4625(5) \\
			& $\Pi^{(1)}$ & 0.470(2) & -0.003(1) & -0.0028(5) & -0.4625(5) \\
			\hline
		\multirow{2}{*}{2} & $\Pi^{(0)}$ & 0.5159(2) & 0.0007(3) & 0.0005(4) & 0.4788(4) \\
			& $\Pi^{(1)}$ & 0.4841(2) & -0.0007(3) & -0.0005(4) & -0.4788(4) \\
			\hline
		\multirow{2}{*}{3} & $\Pi^{(0)}$ & 0.534(1) & 0.003(1) & 0.0029(4) & 0.4600(5) \\
			& $\Pi^{(1)}$ & 0.466(1) & -0.003(1) & -0.0029(4) & -0.4600(5) \\
			\hline
		\multirow{2}{*}{4} & $\Pi^{(0)}$ & 0.5181(6) & 0.001(4) & 0.0004(4) & 0.4417(4) \\
			& $\Pi^{(1)}$ & 0.4819(6) & -0.001(4) & -0.0004(4) & -0.4417(4)
	\end{tabular}
	\caption{Individual measurement.} \label{tab:ibmqx2_ind}
	\end{subtable} \\
	\vspace{5mm}
	\begin{subtable}[t]{0.5\textwidth}
	\begin{tabular}{c | c | c | c | c | c}
		qubit & operator & $a_0$ & $a_1$ & $a_2$ & $a_3$ \\
		\hline
		\multirow{2}{*}{0} & $\Pi^{(0)}$ & 0.544(1) & 0.016(1) & 0.0163(4) & 0.4130(4) \\
			& $\Pi^{(1)}$ & 0.456(1) & -0.016(1) & -0.0163(4) & -0.4130(4) \\
			\hline
		\multirow{2}{*}{1} & $\Pi^{(0)}$ & 0.5199(3) & 0.0115(3) & 0.0109(4) & 0.4703(4) \\
			& $\Pi^{(1)}$ & 0.4801(3) & -0.0115(3) & -0.0109(4) & -0.4703(4) \\
			\hline
		\multirow{2}{*}{2} & $\Pi^{(0)}$ & 0.5181(2) & 0.0320(3) & 0.0318(3) & 0.4749(4) \\
			& $\Pi^{(1)}$ & 0.4819(2) & -0.0320(3) & -0.0318(3) & -0.4749(4) \\
			\hline
		\multirow{2}{*}{3} & $\Pi^{(0)}$ & 0.5304(4) & 0.0244(3) & 0.0250(3) & 0.4634(4) \\
			& $\Pi^{(1)}$ & 0.4696(4) & -0.0244(3) & -0.0250(3) & -0.4634(4) \\
			\hline
		\multirow{2}{*}{4} & $\Pi^{(0)}$ & 0.5149(2) & 0.0121(1) & 0.0121(4) & 0.4594(4) \\
			& $\Pi^{(1)}$ & 0.4851(2) & -0.0121(1) & -0.0121(4) & -0.4594(4)
	\end{tabular}
	\caption{Parallel measurement.} \label{tab:ibmqx2_para}
	\end{subtable}
	\caption{Single-qubit detector results for IBM Q 5 Yorktown, (a) measured individually for each physical qubit leaving the other qubits idle; (b) measured for all five qubits in parallel.} \label{tab:ibmqx2}
	\end{table} \vspace{5mm}

\section{QDT for IBM Q 16 Melbourne}

	\begin{figure*}[t]
		\centering
		\includegraphics[width=1\textwidth]{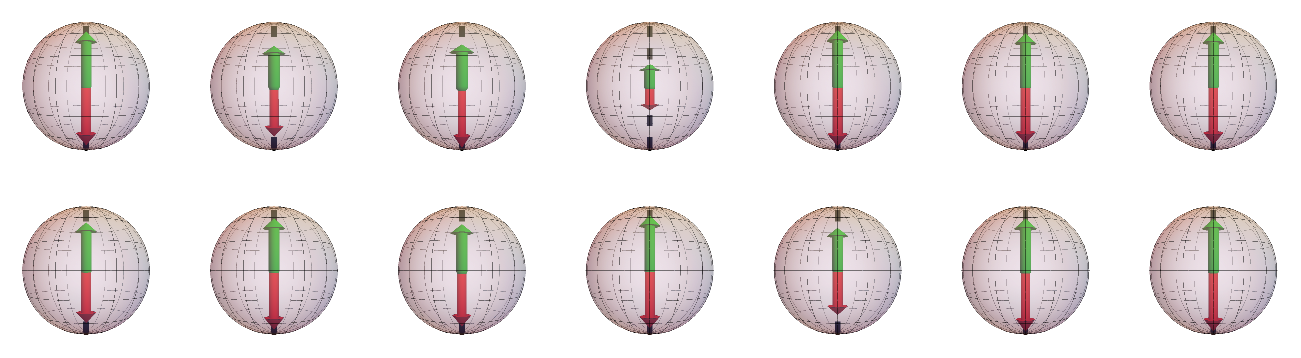}
		\caption{Detector spheres for qubits 0 to 13  of IBM Q 16 Melbourne (with 0 to 6 from left to right on the top row and 7 to 13 on the bottom row).} 
		\label{fig:Qx16}
	\end{figure*}

For completeness, we present the detector tomography on all 14 qubits of the IBM Q 16 Melbourne device. The results are presented in the form of Bloch spheres in Fig.~\ref{fig:Qx16}, done via data taken in parallel for all 14 qubits. As in the other machines, all detectors align pretty much along the vertical z-axis. We notice that the POVMs for qubit 3 have arrows that are relatively shorter than all the rest. Its detectors have the largest imperfection.  We note that one can repeat individual, pairwise, or triple characterization we discussed in the main text, but we do not present them here.

\begin{figure}[h]
		{\centering
		\includegraphics[width=0.45\textwidth]{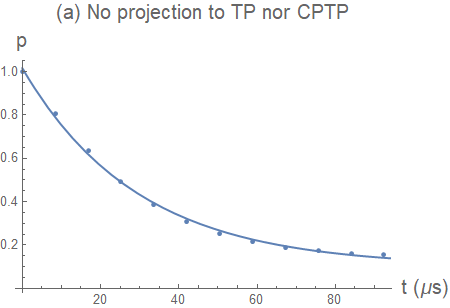} \\
		\vspace{0.5cm}
		\includegraphics[width=0.45\textwidth]{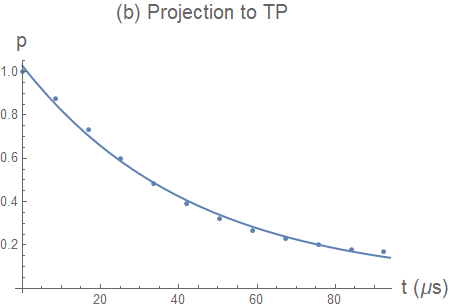} \\
		\vspace{0.5cm}
		\includegraphics[width=0.45\textwidth]{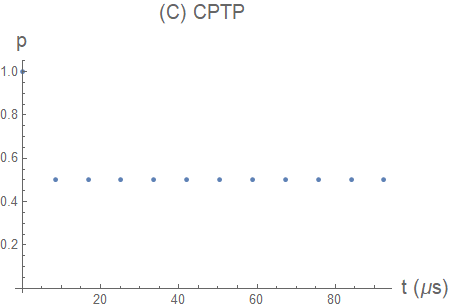}
		}
		\caption{\label{fig:relaxation} Qubit relaxation: probability $p$ of measuring $|1\rangle$ vs. a time $t$, calculated using the identity gate from (a) best GST estimate without imposing TP nor CPTP condition; (b) projecting the identity gate from (a)  to be TP; (c) projecting the identity gate from (a) to be CPTP. }
	\end{figure}

	\section{Relaxation from GST} \label{sec:relaxation}
	
	Here we illustrate the relaxation from GST. Since in our GST implementation we have only used the indentity gate ${G_I}=\openone$ in the germ set, we can examine whether the identity operation, which is essentially letting the qubit idle, can allow us to extract relaxation of a qubit in the excited state $|1\rangle$. Since the circuits for the GST include gate sequence such as $( G_I)^m {G_x}{G_x}$, we already have the data for relaxation. Note that $({G_x})^2=i\sigma_x$ flips $|0\rangle$ to the excited state $|1\rangle$ and the $m$ identity gates represent an idling of $m$ units of gate duration, including gate and buffer times. The sequence measures relaxation. Let us use qubit 0 of ibmqx4 for illustration. Note that each single-qubit gate duration, including the buffer time, is 70$ns$. The identity gate we obtained from GST before imposing TP or CPTP condition, expressed in the Pauli basis, is
	\begin{equation}
		{G_I}=\begin{pmatrix}
		0.998699 & 0.001783& -0.001243  &0.002354\cr
		-0.047276 & 1.044539 &-0.065670 & 0.083989\cr
		0.030418 &-0.068910 &  0.980521 &-0.054349\cr
		0.049455 &-0.093771  &0.056438 & 0.904232
		\end{pmatrix}.
	\end{equation}
	Note that the element $({G_I})_{\alpha,\beta}$ represents the amplitude that $\sigma_\beta$ is mapped to $\sigma_\alpha$ (with $\sigma_0=\openone$), under the idling operation that is supposed to be the indentity gate. This allows us to extract a relaxation time $T_1\approx 29.5 \mu s$. In Fig.~\ref{fig:relaxation}a, we show the probability of obtaining $|1\rangle$ after first applying an ideal $\sigma_x$ gate to $|0\rangle$ and then applying $m$ (=120, 240, \dots, etc.) such discrete identity gates. The curve is an exponential decay fit to the discrete data. Since the obtained idenity gate is not positive, using it to simulate other process, such as $T_2$ decoherence time, we would obtain some probability that is negative, an unphysical result (not shown).

	Projecting the gate to be TP, then the identity gate becomes
	\begin{equation}
		{G_I}=\begin{pmatrix}
		1.    &      0.  &        0.  &        0.        \cr
		-0.047276 & 1.044539 &-0.065670&  0.083989\cr
		0.030418 &-0.068910 & 0.980521 &-0.054348\cr
		0.049455 &-0.093771  &0.056438 & 0.904232
		\end{pmatrix}.
	\end{equation}
	From simulating the relaxation using this identity gate, we extract a relaxation time $T_1\approx 43.2 \mu s$; see Fig.~\ref{fig:relaxation}b. This value differs about 50\% with one obtained earlier without projecting the identity gate to be TP.

	Projecting the gate to be CPTP, then the identity gate becomes
	\begin{equation}
		{G_I}=\begin{pmatrix}
		1. & 0. & 0. & 0.\cr
		0. &  0.9731358 &-0.02961343 &  0.08478270\cr
		0. & -0.03283288 &   0.9433932 & -0.05294062\cr
		0.&-0.08548849 & 0.05312586 & 0.9170383
		\end{pmatrix}.
	\end{equation}
	However, from this matrix, we cannot extract a meaningful relaxation time; see Fig.~\ref{fig:relaxation}c.  Although it is out of the scope of this paper, it will be interesting to see if one can characterize the identity gate (which lets the qubit idle) and obtain its correct CPTP description that can describe relaxation or even dephasing.

\end{document}